\documentclass{article}
\date{} 

\usepackage{graphicx}       
\usepackage{acronym}
\usepackage{amssymb}        
\usepackage[a4paper, margin=2.5cm]{geometry} 
\usepackage{float}

\usepackage{array}
\usepackage{tabularx}
\newcolumntype{Y}{>{\raggedright\arraybackslash}p{0.20\textwidth}}
\usepackage{longtable}
\usepackage{booktabs}
\usepackage{multirow}
\usepackage{tabularx}
\usepackage{appendix}
\usepackage{hyperref}       
\usepackage{xcolor}
\usepackage{todonotes}
\usepackage{xcolor}

\DeclareUnicodeCharacter{202F}{\,}
\usepackage{xcolor}
\usepackage{marginnote}

\makeatletter
\def\@fnsymbol#1{\ensuremath{\ifcase#1\or *\or \dagger\or \ddagger\or 
   \mathsection\or \mathparagraph\or \|\or **\or \dagger\dagger 
   \or \ddagger\ddagger \else\@ctrerr\fi}}
\makeatother

\title{Operationalising AI Regulatory Sandboxes: Activities, Requirements, and Technical Assessment under the EU AI Act}

\author{
  Alessio Buscemi$^{*}$ \and
  Thibault Simonetto$^{+}$ \and
  Daniele Pagani$^{*}$ \and 
  German Castignani$^{*}$ \and
  Maxime Cordy$^{+}$ \and
  Jordi Cabot$^{*}$ \\
  \\
  $^{*}$Luxembourg Institute of Science and Technology (LIST), \texttt{\{name.surname\}@list.lu} \\
  $^{+}$University of Luxembourg, \texttt{\{name.surname\}@uni.lu}
}

\begin{document}

\maketitle

  \acrodef{AI}{Artificial Intelligence}
  \acrodef{FCA}{Financial Conduct Authority} 
  \acrodef{AI Act}{Artificial Intelligence Act}
  \acrodef{GDPR}{General Data Protection Regulation}
  \acrodef{CRA}{Cyber Resilience Act} 
  \acrodef{SME}{Small-medium enterprise}
  \acrodef{GPAI}{General Purpose AI}
  \acrodef{DGA}{Data Governance Act}
  \acrodef{DSA}{Digital Services Act}
  \acrodef{DMA}{Digital Markets Act} 
  \acrodef{AIRS}{AI Regulatory Sandbox}
  \acrodefplural{AIRS}[AIRS]{AI Regulatory Sandboxes}
  \acrodef{AITS}{AI Technical Sandbox}
  \acrodef{EUSAIR}{EU Regulatory Sandboxes for AI}
  \acrodef{LIST}{Luxembourg Institute of Science and Technology} 
  \acrodef{CNPD}{National Data Protection Commission}
  \acrodef{EDIH}{European Digital Innovation Hub} 
  \acrodef{TEF}{Testing and Experimentation Facility}
  \acrodefplural{TEF}[TEFs]{Testing and Experimentation Facilities}
  \acrodef{EuroHPC}{European High-Performance Computing}
  \acrodef{EDIC}{European Digital Infrastructure Consortia}
  \acrodef{AIoD}{AI-on-Demand Platform}
  \acrodef{AI HLEG}{High-Level Expert Group on Artificial Intelligence}
  \acrodef{AI RMF}{AI Risk Management Framework}
  \acrodef{CA}{Competent Authority}
  \acrodefplural{CA}[CAs]{Competent Authorities}
  \acrodef{NIST}{National Institute of Standards and Technology}
  \acrodef{RMF}{Risk Management Framework}
  \acrodef{CEN}{European Committee for Standardization}
  \acrodef{CENELEC}{European Committee for Electrotechnical Standardization}
  \acrodef{JTC}{Joint Technical Committee}
  \acrodef{RAG}{Retrieval Augmented Generation}
  \acrodef{NLP}{Natural Language Processing}
  \acrodef{HPC}{High Performance Computing}
  \acrodef{DSL}{Domain-Specific Language}
  \acrodef{RISE}{Research Institutes of Sweden}
  \acrodef{NFR}{requirement}
  \acrodef{EC}{European Commission}
  \acrodef{OECD}{Organisation for Economic Co-operation and Development}
  \acrodef{API}{Application Programming Interface}
  \acrodef{RE.M.I.}{Regulation Meets Innovation}
  \acrodef{L-AIF}{Luxembourg AI Factory}
  \acrodef{CNPD}{Commission Nationale pour la Protection des Données}
  \acrodef{BESSER}{Building bEtter Smart Software fastER}
  \acrodef{SnT}{Interdisciplinary Centre for Security, Reliability and Trust}
\maketitle

\begin{abstract}
The systematic assessment of AI systems is increasingly vital as these technologies enter high-stakes domains. To address this, the EU's \textit{Artificial Intelligence Act} introduces \acp{AIRS}: supervised environments where AI systems can be tested under the oversight of \acp{CA}, balancing innovation with compliance, particularly for startups and SMEs.
Yet significant challenges remain: assessment methods are fragmented, tests lack standardisation, and feedback loops between developers and regulators are weak.

This paper operationalises the AIRS lifecycle. We map the sandbox journey into 29 concrete activities, from pre-participation guidance through application, preparation, participation, exit, and post-participation monitoring, and we distinguish between a \textit{Core AIRS} centred on regulatory oversight and an \textit{Extended AIRS} that additionally embeds structured technical testing through an \ac{AITS}. From this mapping we derive 15 infrastructural and governance requirements that an AITS must satisfy, each linked to the activities it supports and, for high-risk systems, to the provider obligations set out in Articles~9--15 of the AI Act.

The framework aims to address multiple stakeholders: \acp{CA} gain structured workflows for applying legal obligations; technical experts can integrate robust evaluation methods; and AI providers access a transparent pathway to compliance. We conclude by outlining the \textit{Sandbox Configurator}, an open-source framework intended to instantiate AITS environments from these requirements, and by discussing how a shared technical foundation can support a scalable and innovation-friendly European infrastructure for trustworthy AI governance.
\end{abstract}


\section{Introduction}
\label{sec:introduction}

\ac{AI} systems are increasingly embedded in critical sectors such as healthcare, public administration, finance, and transport, where failures or biases may carry significant consequences for individuals and society at large \cite{bengio2024managing, wang2022artificial}. In response, regulatory efforts are intensifying to ensure that such systems are trustworthy, fair, and transparent \cite{novelli2024taking}. The EU's \ac{AI Act} \cite{eu_ai_act_2024} represents a landmark legislative step in this direction, introducing a risk-based regulatory framework that mandates conformity assessment for high-risk applications.

Among the Act's key instruments are \ac{AIRS}es, or \textit{AI Regulatory Sandboxes}, supervised environments where AI systems can be tested in collaboration with Competent Authorities (\acp{CA}) and technical experts. These sandboxes are designed to promote innovation while ensuring regulatory compliance, particularly for SMEs and startups. All Member States are required to implement operational sandboxes by August 2026. 
The FARI white paper on AI sandboxes \cite{due2024sandboxing} highlights that their effectiveness depends on use-case–driven participation, a balance between regulatory and technical testing, and trust-building among innovators, regulators, civil society, and experts.
The value of controlled testing environments has also gained international recognition. In the United States, the AI Action Plan \cite{aiactionplan} includes provisions for agency-specific sandboxes, overseen by regulators such as the FDA and SEC, to facilitate the safe and accountable development of AI across sectors such as healthcare and finance.

First, AI assessments must be adapted to the context of use, as different domains demand distinct metrics, thresholds, and methods. Second, the growing number of open-source and proprietary tests lacks standardisation, making it laborious to identify, install, and consolidate results across tools. Third, developers and regulators still lack effective channels for feedback, which prevents them from learning from one another and adjusting to risks over time.
To address these challenges, this paper establishes an operational baseline for AIRS implementation. We formalise the sandbox lifecycle into processes and activities, and derive from them the requirements that a technical assessment environment must satisfy in order to support trustworthy, reproducible, and legally traceable evaluation. On this basis, we outline the \textit{Sandbox Configurator}, a modular, open-source framework intended to instantiate customised sandbox environments from such requirements; its concrete design and implementation are reported in a companion paper \cite{configurator2026}.

We organise our contribution as follows:
\begin{itemize}

    \item \textbf{First}, we map the AIRS lifecycle and formalise it into processes and 29 concrete activities, spanning the full journey from pre-participation to post-participation monitoring. This mapping is situated within the broader vision of "regulatory sandboxes à la carte" proposed by EUSAIR~\cite{eusair2025}.
    
    \item \textbf{Second}, we derive from these activities 15 infrastructural and governance requirements for the \ac{AITS}, mapping each requirement to the activities it supports and, for high-risk systems, to the corresponding provider obligations under Articles~9--15 of the AI Act.

    \item \textbf{Third}, we reflect on how such a framework complements regulatory oversight, supports company-led self-assessment, and fosters harmonised sandboxing practices across Europe.
\end{itemize}

The framework targets a broad audience. 
For \acp{CA}, it offers a concrete blueprint to translate legal mandates into structured supervisory workflows. 
For technical experts, it provides a principled way to integrate their methods into evaluative environments that are both robust and compliant. 
AI providers, particularly SMEs, are expected to benefit from a transparent, phased pathway for experimentation, regulator-guided feedback, and increased readiness for market deployment. 
Civil society stakeholders are also actively integrated through the framework's emphasis on transparency, explainability, and auditability.

Finally, we outline a forward-looking vision in which a shared technical foundation for sandboxing serves as an enabler of cross-border cooperation. 
By supporting the shared development, adaptation, and governance of sandbox configurations, it facilitates the emergence of interoperable testing environments and coordinated supervisory practices. 
This positions Europe to not only enforce trustworthy AI standards at scale, but also to lead in shaping a model of innovation-friendly, rights-preserving AI governance.

Throughout this paper, including the background discussion in Section~\ref{sec:background} and the mapping of processes and activities in Section~\ref{sec:activities}, our analysis is grounded in our interpretation of the AI Act together with the currently available documentation on regulatory sandboxes. 
At this stage, no official technical guidelines or harmonised standards have yet been issued, which means that both our conceptual framing and the proposed framework inevitably reflect a provisional understanding of the law. 
As the AI Office and standardisation bodies progressively release further guidance and supporting instruments, we anticipate that future versions of this work will revisit, refine, and extend the analysis presented here so as to remain fully aligned with the evolving regulatory landscape.

\section{Background}
\label{sec:background}

This section offers the contextual foundations for understanding the paper's contributions. 

\subsection{Regulatory sandboxes}
\label{sub:regulatory_sabdbox}

The 2008 financial crisis exposed critical weaknesses in financial oversight and led to a global push for more resilient, transparent, and consumer-friendly financial systems. At the same time, a wave of fintech innovation, spanning digital payments, blockchain, and robo-advisory, began challenging traditional regulatory frameworks.  

To balance innovation with stability, the UK \ac{FCA} introduced the world's first regulatory sandbox in 2016 \cite{authority2017regulatory}. It allowed financial companies to qualify and test new products in a controlled environment under regulatory supervision, with temporary relief from certain compliance requirements. This approach enabled faster time-to-market while safeguarding consumers and the financial system. Inspired by the UK model, countries such as Singapore, Australia, and Bahrain quickly followed. By the early 2020s, over 70 jurisdictions had adopted regulatory sandboxes \cite{appaya2020global}, cementing them as a key policy tool for managing financial innovation responsibly. 

More recently, the \ac{OECD} has framed regulatory sandboxes as part of a broader toolkit for adaptive governance \cite{oecd2025toolkit}. According to this perspective, sandboxes are most effective when used in contexts of regulatory uncertainty combined with high innovation potential, where live testing provides insights that cannot easily be obtained through conventional tools such as consultations or pilot projects. The OECD highlights four dimensions that regulators must consider when designing and implementing a sandbox: 

\begin{enumerate}
    \item \textit{Legal mandate and proportionality}, i.e., whether the authority has the power to grant exemptions or waivers, and whether sandbox interventions are the least intrusive means available;  
    \item \textit{Stakeholder engagement}, ensuring innovators, consumers, and civil society are included to avoid capture and to surface societal concerns early;  
    \item \textit{Risk management and monitoring}, requiring clear criteria for entry and exit, predefined safeguards, and ongoing data collection to mitigate consumer and systemic risks;  
    \item \textit{Learning and policy feedback}, as sandboxes should not only enable product testing but also provide evidence for potential regulatory reforms.  
\end{enumerate}

\subsection{AI Regulatory Sandboxes}
\label{sub:ai_regulatory_sandbox}

The \ac{AI Act} is the European Union's landmark legislation designed to regulate the development, deployment, and use of \ac{AI} across Member States. It aims to ensure that AI systems placed on the EU market are safe, respect fundamental rights, and promote trustworthy innovation. Central to the Act is a risk-based regulatory approach, which classifies AI systems into four categories, minimal, limited (referred to as \textit{ transparency obligations}), high, and unacceptable risk, based on their potential to harm health, safety, or fundamental rights. This framework enables proportional obligations, with the strictest requirements applying to high-risk systems, including obligations for transparency, data governance, human oversight, and robustness.

The \ac{AI Act} introduces regulatory sandboxes as controlled environments where AI providers can assess and validate innovative systems under the supervision of a \ac{CA} prior to market deployment. 
These environments serve a dual function: on the one hand, they facilitate technological experimentation by offering developers a supervised setting to evaluate system performance, risks, and legal compliance; on the other, they ensure that such innovation unfolds in alignment with EU values, safeguarding fundamental rights and the public interest across the entire development lifecycle. Crucially, this alignment is informed by the seven principles of Trustworthy AI identified by the \ac{EC}, such as human agency, transparency, and societal well-being, which provide an ethical compass for both technical assessment and regulatory interpretation \cite{eu_trustworthy_ai_2019}. These principles underpin the design of AIRS and guide the evaluation of AI systems beyond strict legal compliance, ensuring that innovation contributes to the broader public good.

Under Article 57 of the AI Act, all Member States are required to establish at least one \ac{AIRS} by 2 August 2026. Participation in these sandboxes is entirely voluntary, and they must be offered free of charge, at least for core activities (see Section \ref{sub:aits} for further details), to startups and SMEs, supporting the EU's goal of democratising AI development and reducing compliance costs for smaller players.
This measure is particularly critical in ensuring that regulatory complexity does not become a barrier to entry for promising but resource-constrained providers.
The Act identifies three types of sandboxes to accommodate different regulatory and operational needs.

\textbf{National transversal Sandboxes}: These are managed by Member States and provide localised support to AI providers and deployers. They are instrumental in promoting national ecosystems of innovation and facilitating the development of AI solutions adapted to domestic socio-economic contexts.

\textbf{National sector-specific Sandboxes}: These sandboxes are tailored to the distinct characteristics of specific application areas (e.g., healthcare, finance, education, law enforcement), and may involve collaboration with sectoral regulators or data protection authorities. They allow for contextualised testing where the interaction between the AI Act and sectoral legislation requires careful calibration.

\textbf{Cross-border Sandboxes}: These are designed to support AI systems operating across multiple Member States. They promote regulatory convergence and consistent supervision across jurisdictions, which is essential for scalable AI applications. It is important to note that cross-border sandboxes can be both transversal and sector-specific.

Participation in \ac{AIRS}es grants AI actors several benefits. Authorities provide regulatory guidance and hands-on support to clarify obligations, interpret ambiguous provisions, and offer iterative feedback on system design, risk management, and conformity pathways. Under certain conditions, and provided fundamental rights are not at risk, temporary derogations may be granted for phased compliance (e.g., documentation or logging duties). Supervised real-world testing enables providers to evaluate systems in operational contexts under regulatory oversight, though liability remains with them. Unlike a conformity assessment, \ac{AIRS}es offer a pre-commercial environment to align systems with regulatory expectations; the \textit{Exit Report} can later serve as supporting evidence during audits. 
Beyond practical utility, \ac{AIRS}es are strategic tools: they empower SMEs and research actors to adopt compliance-by-design, enhance legal certainty for grey-zone technologies, strengthen regulatory foresight, inform the development of standards and certification, and promote cross-border collaboration towards a coherent EU AI market.

In sum, \ac{AIRS}es operationalise the core ambition of the AI Act: to ensure that technological advancement proceeds hand in hand with human-centric values, institutional trust, and legal predictability.

\subsection{Supporting technical assessment within AIRS}
\label{sub:aits}

Across Europe, a growing number of initiatives are labeled as "AI sandboxes'', e.g. \cite{list2024aisandbox, unicc_sandbox_2025}, but it is essential to distinguish between an \ac{AITS} and an \ac{AIRS}, as they serve fundamentally different purposes. While both aim to promote the development of trustworthy AI, an \ac{AITS} provides a technical environment for experimentation, evaluation, and compliance-by-design outside the formal legal framework. In contrast, an \ac{AIRS} facilitates direct collaboration with \acp{CA}, offering supervised settings to interpret and apply regulatory requirements under the EU AI Act. An AITS can be decomposed into two interlinked layers: a \textbf{Development Sandbox} and \textbf{Assessment Sandbox}.

A development sandbox enables early-stage experimentation by embedding compliance principles such as transparency, explainability, and robustness directly into the design phase of AI systems.
An assessment sandbox supports post-hoc technical evaluations, covering properties such as accuracy, biases, robustness, transparency and energy efficiency, at the critical juncture before deployment or market entry.
While the AI Act does not explicitly use the term \textit{Assessment Sandbox}, Article 58(2)(i) mandates that \ac{AIRS}es should:

\begin{quote}
"facilitate the development of tools and infrastructure for testing, benchmarking, assessing and explaining dimensions of AI systems relevant for regulatory learning, such as accuracy, robustness and cybersecurity, as well as measures to mitigate risks to fundamental rights and society at large.''
\end{quote}

This provision makes clear that \ac{AIRS} are not confined to legal supervision, but must also integrate 
\textbf{technical evaluations} that contribute to regulatory learning and inform risk mitigation. At the same time, since participation in the \ac{AIRS} is entirely voluntary, its format should adapt to the level of commitment and the needs of the participant. 
In some cases, the AI provider may already have assessment results, conducted internally or by third parties, that can serve as the basis for regulatory guidance. 
In other cases, the provider may wish to conduct the necessary testing during the sandbox journey itself, requiring closer integration of technical expertise.

To reflect this diversity of participant pathways, we introduce a practical, \textbf{participant-oriented distinction} 
between two modalities of engaging with the \ac{AIRS}:
\begin{itemize}
    \item \textbf{Core AIRS}, centred on legal and procedural oversight, particularly regarding risk classification, conformity pathways, and regulatory guidance;
    \item \textbf{Extended AIRS}, which build on the Core by embedding structured technical testing and hands-on evaluation of robustness, accuracy, bias, cybersecurity, and other risk dimensions.
\end{itemize}

This differentiation is analytical only: it does not imply that two separate \ac{AIRS} settings exist in law. 
Rather, it clarifies the different journeys available to participants depending on whether they seek regulatory guidance alone or combined with technical assessment. 
A comparable usage has already been acknowledged by EUSAIR, which refers to \textit{AIRS 1.0} (aligned with our Core AIRS) and \textit{AIRS 2.0} (aligned with our Extended AIRS)~\cite{eusair2025}. 
The examples of AIRS journeys presented in Section~\ref{sec:examples} illustrate how participants may engage with the AIRS in these two different ways. 
This distinction also helps to anticipate the kind of resources that regulators may need: in a Core AIRS the regulator alone is sufficient, whereas in an Extended AIRS additional support from technical experts, whether hired internally or subcontracted as independent evaluators, becomes necessary. 
Section~\ref{sub:stakeholders} further clarifies the identity and roles of these technical experts.

In practice, the AIRS constitutes a single environment that flexibly adapts to participants' needs, generating evidence-based insights to support regulatory decision-making, and providing assessment sandbox capabilities whenever required by the participant.

\subsection{Stakeholders and Actors of AIRS}
\label{sub:stakeholders}

In Table \ref{tab:airs-stakeholders}, we summarise the key stakeholders and actors involved in the \ac{AIRS} framework. 
We distinguish between \textit{stakeholders}, who are directly benefiting from or impacted by AIRS, and \textit{actors}, who provide operational, technical, or policy support. 
Each category plays a distinct role: \textit{AI providers} engage as the primary beneficiaries of sandboxing activities; \textit{national \acp{CA}} act as coordinators and guarantors of legal oversight; \textit{technical experts} contribute independent methodologies and testing capabilities; the \textit{AI Office} connects sandbox experiences to policy learning at the EU level; \textit{European initiatives} ensure alignment with broader infrastructures; and finally, \textit{civil society} actors safeguard transparency, fairness, and legitimacy. 
For the category of European Initiatives, we draw specifically on the EUSAIR report~\cite{eusair2025}, which situates AIRS within the broader EU AI innovation ecosystem.

\begin{table*}[h]
\centering
\caption{Stakeholders and Actors of AIRSes}
\label{tab:airs-stakeholders}
\begin{tabular}{p{1.8cm} p{1.8cm} p{11.5cm}}
\hline
\textbf{Type} & \textbf{Category} & \textbf{Description and Role in AIRS} \\
\hline
Stakeholder & AI Providers & 
\textit{SMEs and Startups}: Seek legal certainty, faster time-to-market, and credibility; benefit from tailored regulatory guidance, risk management, and documentation support, while confidentiality and trust-building are key. \newline
\textit{Companies with Disruptive AI Systems}: Operate in regulatory grey zones; sandboxing can provide experimentation space, independent technical assessment, and co-development of oversight practices. \newline
\textit{Large Companies}: Use AIRS for independent assurance, validation of compliance processes, and demonstration of responsible innovation. \newline
\textit{Public Sector Organisations}: Explore AI for citizen services and efficiency; sandboxing can ensure alignment with rights, transparency, and democratic accountability. \\
\hline
Stakeholder & National \acp{CA} & 
Central coordinators of \ac{AIRS}es; review applications, define legal scope, monitor compliance. Provide interpretative guidance rather than formal conformity assessments. Balance neutrality, transparency, and support while gathering insights for EU-level policy. \\
\hline
Actor & Technical Experts & 
Independent evaluators (e.g., standardisation bodies, notified bodies, research centres, private firms) contributing domain-specific testing, advisory assessments, and methodology development. Extend regulatory frontier for emerging technologies without issuing binding certifications. \\
\hline
Actor & AI Office & 
Receives feedback from \ac{AIRS}es to inform evidence-based policymaking, anticipate bottlenecks, and adapt regulations.\\
\hline
Actor & European Initiatives & 
\textit{\acp{EDIH}}: Act as entry points for SMEs and startups; triage needs, provide training, channel funding, and perform risk assessments \cite{eu_edihs}. \newline
\textit{AI-on-Demand Platform}: Offer datasets and tools, support experimentation \cite{aiod_platform_2023}. \newline
\textit{Data Spaces}: Supply datasets for training, validation, and evaluation \cite{dataspace}. \newline
\textit{\acp{TEF}}: Enable sector-specific real-world testing \cite{agrifoodtef2025,tefhealth2025,aimatters2025,citcom_ai_2025}. \newline
\textit{EuroHPC \& AI Factories}: Deliver large-scale training and testing, provide sector-specific services \cite{eurohpc,eu_ai_factories}. \\
\hline
Actor & Civil Society & 
\textit{Consumer organisations, advocacy groups, ethicists, journalists, rights associations}: Provide independent oversight, raise societal concerns, ensure inclusiveness and legitimacy. Enhance transparency, fairness, and alignment with democratic values. \\
\hline
\end{tabular}
\end{table*}

\section{Mapping AIRS Processes and Activities}
\label{sec:activities}

\begin{figure*}
    \centering
    \includegraphics[width=1\textwidth]{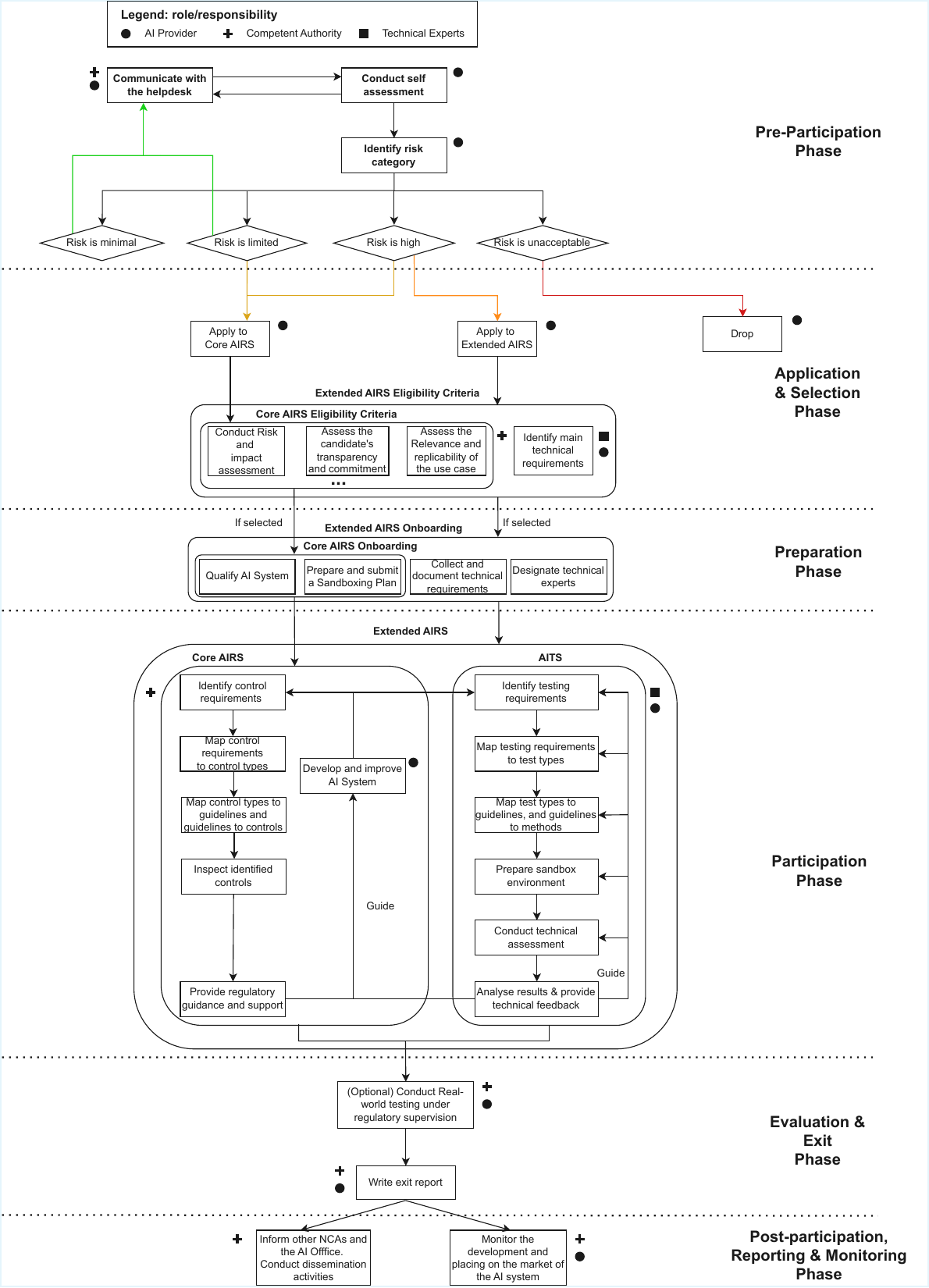}
    \caption{AI Regulatory Sandbox Journey}
    \label{fig:sandbox_framework}
\end{figure*}

Figure~\ref{fig:sandbox_framework} offers a structured overview of the \ac{AIRS} journey, showing how collaboration between regulators and technical experts can evolve throughout the full lifecycle. The journey is divided into phases, each corresponding to specific responsibilities, decision points, and objectives set out in the AI Act. These phases mirror the division proposed by EUSAIR~\cite{eusair2025} and are aligned with steps identified in the OECD regulatory sandboxes toolkit \cite{oecd2025toolkit}.

The purpose of this section is to formalise the sandbox journey into a set of processes and activities. This formalisation does not only provide clarity on how the phases unfold, but also serves a practical role: it enables the systematic collection of requirements for the technical assessment environment introduced in Section~\ref{sec:building_aits}. Each phase is therefore summarised, and its main activities are presented in a dedicated table, ensuring that the operational logic of AIRS is directly mapped into actionable inputs for the design of the \ac{AITS}.

\subsection{Pre-Participation Phase Activities}
\label{sub:preparatory_phase}

Before entering the sandbox, prospective participants should complete a preparatory phase to ensure alignment with its scope, regulatory expectations, and legal applicability under the AI Act. This stage involves informal support through a helpdesk, an initial risk classification of the AI system, and the submission of a reasoned request for AIRS participation.

During the Pre-Participation phase, providers are expected to carry out a self-assessment to determine the risk category of their AI system. The helpdesk serves as the first point of contact in this process by offering guidance and advice that can support providers in performing the self-assessment and in understanding the available participation paths. Beyond this entry role, the helpdesk also responds to broader inquiries on the interpretation and implementation of the AI Act, in line with Article 70(8), which empowers national \acp{CA} to provide guidance and advice, particularly to SMEs and start-ups.

If a system is classified as minimal risk, the provider continues dialogue with the helpdesk without entering AIRS. Limited-risk systems allow providers to either remain engaged with the helpdesk or submit an application to the Core AIRS. High-risk systems should normally lead to an application to the Extended AIRS, whereas systems involving prohibited practices require providers to cease development or reduce the level of risk until the system falls within a permissible category.

Table~\ref{tab:preparatory_phase_activities} outlines the key tasks taken into consideration to ensure effective early engagement with prospect participants.

\begin{table}[H]
\centering
\caption{Activities for the Pre-participation Phase}
\label{tab:preparatory_phase_activities}
\begin{tabular}{p{3.5cm} | p{9.5cm} | p{2.5cm}}
\toprule
\textbf{Activity} & \textbf{Description} & \textbf{Responsible} \\
\midrule
\textbf{A1: Providing Helpdesk Guidance} & 
Set up a helpdesk as the first point of contact for prospective participants. Offer informal, non-binding guidance on AI Act applicability, system categorisation, and procedural expectations. Coordinate with legal or sectoral experts for complex enquiries when needed. & 
\ac{CA} \\
\hline
\textbf{A2: Supporting Preliminary Risk Classification} & 
Develop and distribute a structured questionnaire to help providers perform a self-assessment of their AI system's risk category in accordance with the AI Act. & 
\ac{CA} \\
\bottomrule
\end{tabular}
\end{table}

\subsection{Application \& Selection Phase Activities}
\label{sub:application_phase}

As outlined in Section~\ref{sub:preparatory_phase}, AI systems classified as high risk are generally advised to engage with Extended AIRS, while those deemed limited risk should opt for Core AIRS. 
This distinction reflects the varying needs of AI providers, some may seek technical support and iterative testing, while others may be primarily interested in regulatory guidance.
Table~\ref{tab:application_phase_activities} presents the key activities to be fulfilled during the Application \& Selection phase, ensuring a consistent and effective onboarding process for all participants.

\begin{table}[H]
\centering
\caption{Activities for the Application \& Selection Phase}
\label{tab:application_phase_activities}
\begin{tabular}{p{3.5cm} | p{9.5cm} | p{2.5cm}}
\toprule
\textbf{Activity} & \textbf{Description} & \textbf{Responsible} \\
\midrule
\textbf{A3: Managing Application Submissions} & 
Design and operate a structured application process for AIRS participation, collecting detailed information on the AI system's purpose, intended users, potential impacts on fundamental rights, existing safeguards, and the rationale for joining the sandbox. & 
\ac{CA}\\
\hline
\textbf{A4: Conducting Evaluation and Admission} & 
Implement a transparent, criteria-based evaluation procedure to assess the suitability of applicants. Evaluate their innovation potential, readiness for controlled testing, understanding of legal obligations, and capacity to document and share outcomes. Prioritise projects with societal relevance, rights implications, and adequate resourcing. & 
\ac{CA}, with the support of Technical Experts when needed.\\
\bottomrule
\end{tabular}
\end{table}

\subsection{Preparation Phase Activities}
\label{sub:onboarding_activities}

The Preparation phase is intended to validate that the system's declared scope, intended use, and technical specifications remain consistent with its regulatory designation. 
This step ensures a reliable foundation for subsequent supervision and eventual testing within the sandbox. 

\ac{CA} are responsible for implementing a structured onboarding process that enables accurate qualification, facilitates planning, and ensures the collection of relevant technical information. 
In cases where Extended AIRS is foreseen, this phase should also initiate the identification and engagement of appropriate technical experts to support testing activities. 
Although not all AI systems participating in AIRS will fall under the \textit{high-risk} category, the Preparation phase reflects the first steps of the risk management logic set out in Article~9 of the AI Act. 
For high-risk systems, it contributes directly to risk identification, analysis, and evaluation by clarifying the system's purpose, operational boundaries, and potential impact on fundamental rights or safety. 
For lower-risk systems, these activities still provide valuable structure, transparency, and traceability, thereby fostering regulatory learning and comparability across cases.
Table~\ref{tab:onboarding_activities} outlines the key procedural elements that must be established to operationalise this phase effectively.

\begin{table}[H]
\centering
\caption{Activities for the Preparation Phase}
\label{tab:onboarding_activities}
\begin{tabular}{p{3.5cm} | p{9.5cm} | p{2.5cm}}
\toprule
\textbf{Activity} & \textbf{Description} & \textbf{Responsible} \\
\midrule
\textbf{A5: Qualifying the AI System} &
Conduct a process to verify that the AI system's declared purpose, intended use, and deployment context are consistent with its initial regulatory classification, ensuring eligibility and relevance for sandbox participation. & 
\ac{CA} \\
\hline
\textbf{A6: Submitting the Sandbox Plan} &
Request and review a detailed sandbox engagement plan from participants, outlining their testing scope, participation objectives, key regulatory or technical challenges, and projected timeline. This plan will guide mutual expectations and the sandbox workflow. & 
\ac{CA} \\
\hline
\textbf{A7: Appointing Technical Experts} &
In Extended AIRS, initiate early coordination to identify and engage appropriate technical experts, either internal, external, or affiliated with research/testing facilities, responsible for executing the sandbox assessment. & 
\ac{CA} \\
\hline
\textbf{A8: Documenting Technical Characteristics} &
For Extended AIRS, collect and structure technical information on the AI system's architecture, data flows, interaction modes, deployment constraints, and operational boundaries to establish a sound basis for evaluation. & 
AI Provider (documentation), Technical Experts (validation) \\
\bottomrule
\end{tabular}
\end{table}

\subsection{Participation Phase Activities}
\label{sub:participation}

The participation phase represents the core stage of AIRS. 
As discussed in Section~\ref{sub:aits}, the distinction between Core AIRS and Extended AIRS is purely analytical and reflects the participant's perspective on how they may engage with the single comprehensive sandbox defined by the AI Act. 
We therefore distinguish the activities of this phase between a Core AIRS, centred on regulatory guidance, and an Extended AIRS, which additionally incorporates technical testing.
Before detailing the specific processes and activities in Sections~\ref{sub:core_airs} and \ref{sub:extended_activities}, we first outline the assessment dimensions that should be taken into consideration.

\subsubsection{Dimensions of Assessment in AIRS}
\label{sub:layered}

The analysis of an AI system should be structured along \textbf{three core dimensions} \cite{nist_ai_rmf_2023, iso_iec_42001_2023}: 
1) \textbf{Data and Models}: Covers data provisioning and composition, as well as the design of the model architecture and its integration within the system; 
2) \textbf{Processes}: Encompasses development practices, lifecycle management, and governance mechanisms, including risk and quality management; 
3) \textbf{Final Product}: Focuses on system behavior, user interaction, and broader societal or environmental impacts. 
These dimensions also enable a precise distinction between two complementary oversight mechanisms within the AI Regulatory Sandbox: \textit{controls} and \textit{testing}, each reflecting a distinct role in the supervision process \cite{oecd2025toolkit}. 
For high-risk systems, this distinction aligns with the obligations set out in Articles~9–15 of the AI Act, which jointly require ex-ante safeguards (e.g., risk management, data governance, documentation, transparency, and oversight) and ex-post verification (e.g., monitoring, robustness, accuracy, and cybersecurity). 
For other systems, while these provisions do not apply directly, adopting the same logic still strengthens regulatory learning, comparability, and accountability.

Controls refer to \textit{ex-ante} safeguards, governance practices, and procedural mechanisms that are expected to be implemented by the provider and subsequently verified by the \ac{CA} during the sandbox engagement. 
Controls primarily concern the \textit{Processes} and \textit{Data and Models} dimensions, and may include risk management procedures, data governance policies, documentation protocols, version control practices, and other measures that ensure regulatory alignment throughout the AI system's lifecycle. 
The role of the \ac{CA} is to inspect the presence, adequacy, and coherence of these controls.

Testing, on the other hand, consists of \textit{ex-post} empirical evaluations conducted by technical experts, typically external to the \ac{CA}, to assess the system's performance, behavior, or statistical properties under specific conditions. 
Testing may be applied to both the \textit{Final Product} and the \textit{Data and Models} dimensions. 
For the final product, testing focuses on \textit{behavioral evaluation}, such as verifying output consistency, robustness, or system-level behavior under edge cases. 
For data and models, it involves \textit{statistical evaluation}, such as assessing data distribution across sensitive variables, performance variation across subpopulations, or the presence of unintended bias or drift.

This distinction clarifies the complementary functions within the AIRS: the \ac{CA} verifies the existence and suitability of procedural controls, while qualified technical experts should be mobilised under Extended AIRS to carry out targeted testing.

\subsubsection{Core AIRS Activities}
\label{sub:core_airs}

The Core AIRS is designed around the supervisory responsibilities of the \ac{CA}, with a focus on verifying and reinforcing the internal controls that AI providers must establish to comply with the \ac{AI} Act. These controls are evaluated along the dimensions of \textit{Processes} and \textit{Data and Models}, as introduced in Section~\ref{sub:layered}. They include governance policies, documentation practices, oversight mechanisms, and risk mitigation procedures that should be in place prior to market deployment.

Unlike formal conformity assessments, the Core AIRS does not issue binding certifications. Instead, it promotes a compliance-by-design approach, where the provider receives structured support and early guidance from the regulator. Controls validated through this pathway may later serve as traceable evidence in conformity assessments or audits.
The supervisory role of the \ac{CA} in this context has three primary objectives:

\begin{enumerate}
\item \textbf{Identify the controls that are required} for the specific AI system, based on its intended purpose, risk profile, and deployment context;
\item \textbf{Inspect the controls already implemented} by the provider, evaluating their adequacy, documentation, and coherence with declared system characteristics;
\item \textbf{Provide guidance to improve or complement the control framework}, ensuring that missing, weak, or insufficient controls are addressed and that the system progressively aligns with regulatory expectations.
\end{enumerate}

To operationalise this supervision model, a structured process is needed to trace the translation of abstract legal obligations into concrete implementation steps. Table~\ref{tab:core_airs_control_activities} outlines the key procedural activities that support this control-oriented approach in the Core AIRS.

\begin{table}[H]
\centering
\caption{Activities for Control-Oriented Supervision in Core AIRS}
\label{tab:core_airs_control_activities}
\begin{tabular}{p{3.5cm} | p{8.5cm} | p{2.5cm}}
\toprule
\textbf{Activity} & \textbf{Description} & \textbf{Responsible} \\
\midrule
\textbf{A9: Determining Relevant Control Activities} &
Analyse the declared characteristics of the AI system to identify applicable control activities. These define the safeguards needed to address legal, ethical, and risk-related concerns. & 
\ac{CA} \\
\hline
\textbf{A10: Translating Control Activities into Control Types} &
Convert each control requirement into one or more specific control types, such as traceability mechanisms, human oversight protocols, or dataset versioning procedures. & 
\ac{CA} \\
\hline
\textbf{A11: Aligning Control Types with Guidelines and Operational Measures} &
Map control types to relevant implementation standards, guidelines, or best practices to support providers in translating abstract controls into concrete and operational safeguards. & 
\ac{CA} \\
\hline
\textbf{A12: Reviewing Implemented Controls} &
Inspect the controls implemented by the provider to verify their presence and adequacy in light of the system's purpose and risk profile. This includes documentation review and process inspection, excluding technical testing. & 
\ac{CA} \\
\hline
\textbf{A13: Delivering Regulatory Feedback and Support} &
Provide constructive guidance to help providers improve their control setups and align more closely with regulatory expectations, indicating where safeguards are missing, insufficiently documented, or inconsistent with the declared system characteristics. & 
\ac{CA} \\
\hline
\textbf{A14: Facilitating Shared Traceability of Controls} &
Implement shared tools or collaborative processes that allow both the CA and provider to document, monitor, and update the status of controls throughout the sandbox process. & 
\ac{CA} and AI Provider\\
\bottomrule
\end{tabular}
\end{table}

\subsubsection{Extended AIRS Activities}
\label{sub:extended_activities}

The Extended AIRS includes a dedicated technical evaluation process, carried out through the AITS, which complements legal supervision by enabling testing that is controlled, reproducible, and aligned with established standards and best practices. This environment facilitates the operationalisation of regulatory principles into concrete, measurable evaluation tasks and offers a structured framework for experimentation. To guide this process, we adopt the \ac{RISE} methodology for translating AI Act requirements into structured testing procedures \cite{mowla2024ai}.
The technical assessment proceeds through six sequential stages, beginning with the identification of testing requirements and culminating in the delivery of feedback. Each stage is designed to support a transparent, interoperable, and methodologically robust process that fosters regulatory learning and informs eventual conformity assessment.

\begin{table}[H]
\centering
\caption{Activities for Extended AIRS Implementation}
\label{tab:extended_airs_activities}
\begin{tabular}{p{3.5cm} | p{8.5cm} | p{2.5cm}}
\toprule
\textbf{Activity} & \textbf{Description} & \textbf{Responsible} \\
\midrule
\textbf{A15: Defining Testing Objectives} &
Identify the key aspects of the AI system to evaluate, such as robustness, fairness, transparency, or performance, based on its intended purpose, risk profile, and regulatory obligations. & 
AI Provider and Technical Experts \\
\hline
\textbf{A16: Matching Objectives with Test Types} &
Translate each evaluation objective into suitable test categories (e.g., bias detection, adversarial robustness, performance under constraints) to ensure technical feasibility and alignment with compliance goals. & 
Technical Experts, coordinated by \ac{CA} \\
\hline
\textbf{A17: Linking Tests to Guidelines and Methods} &
Associate each test type with relevant standards, high-quality scientific literature, or best practices, and map them to concrete testing methods and tools to ensure methodological soundness and comparability. & 
Technical Experts \\
\hline
\textbf{A18: Setting Up the Sandbox Environment} &
Build a technical testing setup tailored to the use case, incorporating data, pipelines, dashboards, and infrastructure to ensure traceability, reproducibility, and sector-specific relevance. & 
Technical Experts or AI Provider guided by Technical Experts\\
\hline
\textbf{A19: Executing the Technical Evaluation} &
Run the planned tests within the sandbox environment, ensuring all procedures are traceable and the results are clearly interpretable through adequate visualisation and reporting mechanisms. & 
Technical Experts or AI Provider guided by Technical Experts\\
\hline
\textbf{A20: Synthesising Results and Formulating Feedback} &
Summarise outcomes in a structured technical feedback report for collaborative review with the \ac{CA}, providing actionable insights for system improvement and contributing to conformity documentation and regulatory learning. & 
Technical Experts and AI Provider\\
\bottomrule
\end{tabular}
\end{table}

\subsection{Evaluation \& Exit Phase Activities}
\label{label:after_sandboxing}

Once the main sandboxing phase concludes, participants enter a post-sandbox phase intended to consolidate findings, support regulatory transparency, and demonstrate compliance readiness. This phase includes the submission of an exit report. This report may also serve as part of the written documentation required for future conformity assessment, such as technical reports, control records, and evidence of risk mitigation. While AIRS participation does not constitute a formal conformity assessment, it can significantly support providers in preparing the necessary materials and demonstrating compliance-by-design.

Table~\ref{tab:after_sandboxing} summarises the activities of these phases.

\begin{table}[H]
\centering
\caption{Activities for Evaluation \& Exit Phase}
\label{tab:after_sandboxing}
\begin{tabular}{p{3.5cm} | p{8.5cm} | p{2.5cm}}
\toprule
\textbf{Activity} & \textbf{Description} & \textbf{Responsible} \\
\midrule
\textbf{A21: Preparing Documentation for Real-World Testing} &
Support participants who seek authorisation for real-world testing under the CA's supervision by ensuring that all relevant documentation and insights on the AI system are compiled, assessed, and made available to inform the CA's decision-making. & 
AI Providers (documentation), \ac{CA} (assessment) \\
\hline
\textbf{A22: Compiling and Submitting the Exit Report} &
Coordinate the development of a comprehensive exit report with the participant, summarising technical progress, regulatory findings, evidence of risk mitigation, and remaining challenges. Ensure the report is formatted for use in the conformity assessment process as part of the technical documentation. & 
AI Providers (reporting), \ac{CA} (review) \\
\bottomrule
\end{tabular}
\end{table}

\subsection{Post-Participation, Reporting \& Monitoring Phase Activities}
\label{sub:post_participation_activities}

To ensure that the benefits of sandboxing extend beyond the active participation phase, AIRS should embed mechanisms for structured reporting and long-term monitoring. These activities promote transparency, cross-border consistency, and continuous regulatory learning. Table~\ref{tab:sandbox_post_participation_activities} outlines two key activities.
A central element of the post-sandbox phase is the dissemination of insights. Participation in AIRS should not confer exclusive advantages or distort competition; instead, selected findings should be shared through EU-wide platforms, cooperation with the AI Office, and structured reporting. Such dissemination helps level the playing field, promotes regulatory coherence, and strengthens collective capacity for responsible innovation.

\begin{table}[H]
\centering
\caption{Post-Participation, Reporting \& Monitoring Phase Activities}
\label{tab:sandbox_post_participation_activities}
\begin{tabular}{p{3.5cm} | p{8.5cm} | p{2.5cm}}
\toprule
\textbf{Activity} & \textbf{Description} & \textbf{Responsible} \\
\midrule
\textbf{A23: Sharing Information with Other National CAs and the AI Office. Conduct dissemination activities for the broader public.} & 
Facilitate the dissemination of non-confidential findings from the sandbox, in line with Article 57(8), by transmitting the exit report and other outcomes to the AI Office and, where relevant, publishing them on a European platform. Establishing formal channels for sharing lessons learned and emerging regulatory insights with other \acp{CA} ensures transparency, prevents competitive distortions, and promotes regulatory coherence, anticipatory governance, and harmonisation of practices across Member States. & 
\ac{CA} \\
\hline
\textbf{A24: Monitoring the development and placing on the market of the AI system} & 
Implement follow-up mechanisms to track the evolution of AI providers and their systems once they exit the sandbox. Monitoring may include compliance updates, longitudinal performance checks, ensuring continued alignment with the AI Act. & 
\ac{CA} (monitoring), AI Providers (reporting) \\
\bottomrule
\end{tabular}
\end{table}

\subsection{Ecosystem Integration Activities}
\label{sub:ecosystem_activities}

Outside of the activities that strictly derive from the AIRS phases, we have also identified a set of complementary activities that we consider crucial for the overall success of the sandboxes. To fully leverage the European ecosystem supporting trustworthy AI, the AIRS must be designed to interface with existing infrastructures in a modular, interoperable, and scalable way. Table~\ref{tab:sandbox_ecosystem_activities} outlines the key properties that an AIRS should fulfil in order to enable seamless collaboration across the AI lifecycle.

\begin{table}[H]
\centering
\caption{Core activities enabling AIRS sandbox integration with EU digital and AI infrastructures.}
\label{tab:sandbox_ecosystem_activities}
\begin{tabular}{p{3.5cm} | p{8.5cm} | p{2.5cm}}
\toprule
\textbf{Activity} & \textbf{Description} & \textbf{Responsible} \\
\midrule
\textbf{A25: Enabling Multi-Entry Access} & 
Facilitate access for diverse actors through different infrastructures by implementing lightweight onboarding mechanisms, developing preparatory templates, and providing eligibility checklists tailored to early-stage participants. & 
\ac{CA} (design), Technical Experts (develop) \\
\hline
\textbf{A26: Developing Interoperable Interfaces} & 
Design and implement standardised \acp{API} and metadata schemas within the AITS to allow smooth integration of data and tools with TEFs and AI Factories, ensuring seamless bidirectional sharing of experiments, logs, and results. & 
Technical Experts \\
\hline
\textbf{A27: Supporting Federated Execution} & 
Enable distributed execution of tasks across infrastructures such as TEFs and AI Factories, including training and evaluation activities, while maintaining centralised traceability and compliance logging. & 
Technical Experts, coordinated by \acp{CA} and EU ecosystem \\
\hline
\textbf{A28: Embedding Ecosystem-Aware Governance} & 
Integrate governance mechanisms into the AITS that identify and assign roles to external entities for workflow orchestration, experiment validation, and structured feedback within collaborative AI ecosystems. & 
Technical Experts, guided by the \ac{CA} \\
\hline
\textbf{A29: Aligning Documentation Standards} & 
Adopt and apply common documentation formats (e.g., audit trails, data logs, exit reports) compatible with TEFs and AI Factories to promote validation, reuse, and consistency in evaluation and reporting processes. & 
\ac{CA} and Technical Experts \\
\bottomrule
\end{tabular}
\end{table}

\section{Requirements for Effective AITSs in Extended AIRS}
\label{sec:building_aits}

This section outlines the Requirements that an \ac{AITS} must meet to serve as an effective enabler of the Extended \ac{AIRS}, and sketches how these requirements can be operationalised through a configurable framework.

\subsection{AITS Requirements}
\label{sub:aits_nonfunctional}

The requirements of an \ac{AITS} define the architectural, operational, and governance-level properties that ensure its adaptability, trustworthiness, and interoperability across regulatory domains. 
These do not pertain to specific tests or functionalities, but rather to the foundational capabilities that enable consistent and effective sandboxing. 
For high-risk AI systems, several of these requirements directly support the provider obligations set out in Articles~9–15 of the AI Act, including risk management, data governance, technical documentation, record-keeping, transparency, human oversight, robustness, accuracy, and cybersecurity.
For other systems, while these articles do not apply, the same requirements nonetheless provide value by fostering regulatory learning, comparability, and transparency, thereby reinforcing the broader governance role of AIRS.

Table~\ref{tab:aits_requirements} provides a structured overview of these requirements. 
Each requirement is accompanied by a brief description and is mapped to one or more AIRS activities (Section~\ref{sec:activities}), as well as the specific AI Act articles it touches where relevant. 

\begin{longtable}{p{2.8cm} p{6.6cm} p{2.2cm} p{3.2cm}}
\caption{Requirements for the AITS and their mapping to AIRS activities and AI Act articles.}
\label{tab:aits_requirements}\\
\hline
\textbf{Requirement} & \textbf{Description} & \textbf{Related AIRS Activities} & \textbf{Relevant AI Act Articles (High-Risk Only)} \\
\hline
\endfirsthead

\multicolumn{4}{c}{{\tablename\ \thetable{} -- continued from previous page}} \\
\hline
\textbf{Requirement} & \textbf{Description} & \textbf{Related AIRS Activities} & \textbf{Relevant AI Act Articles (High-Risk Only)} \\
\hline
\endhead

\hline
\multicolumn{4}{r}{{Continued on next page}} \\
\endfoot

\hline
\endlastfoot

\textbf{R1: Customisability \& Modularity} & Provide "à la carte" configuration of tests, metrics, and pipelines so every sandbox engagement is precisely tailored. & A15–A18, A25, A27 & Art.~9 (Risk Mgmt), Art.~15 (Accuracy - appropriate testing)\\
\hline
\textbf{R2: Compatibility with External Catalogue of Assessment Solutions} & Ensure compatibility with external catalogues of assessment tools, benchmarks, and datasets. & A15–A18, A25, A26 & Art.~10 (Data Governance - dataset quality), Art.~13 (Transparency), Art.~15 (Accuracy - appropriate testing)\\
\hline
\textbf{R3: Catalogue of Experts} & Maintain a registry of accredited technical experts able to operate the assessment solutions referenced in R2, interpret their results, and provide advisory support to AI Innovators. & A7, A18–A20 & Art.~14 (Human Oversight - qualified personnel) \\
\hline
\textbf{R4: Visual Pipelines} & Provide low-code, drag-and-drop interfaces for composing, orchestrating, and monitoring test pipelines. & A18, A19 & Art.~11 (Documentation), Art.~13 (Transparency - understandable processes)\\
\hline
\textbf{R5: Tailored Dashboards} & Offer role-specific, real-time dashboards with drill-down and export options. & A18–A20, A22 & Art.~13 (Transparency - accessible information), Art.~14 (Human Oversight - monitoring capabilities)\\
\hline
\textbf{R6: Open-Source Core} & Release the core codebase under a permissive licence to maximise transparency, trust, and community scrutiny. & A14, A17, A25, A29 & Art.~11 (Documentation), Art.~13 (Transparency - algorithmic transparency)\\
\hline
\textbf{R7: Plug-in Architecture} & Expose stable APIs that enable third parties to add sector-specific tests, visualisers, and data connectors. & A16–A18, A25–A27 & Art.~15 (Accuracy - domain-specific validation) \\
\hline
\textbf{R8: Deployment Portability} & Ensure that the AITS can run seamlessly on-premises, in sovereign clouds, or across federated TEF nodes while producing identical artefacts and results. & A18, A21, A25–A27 & Art.~15 (Robustness - consistent performance) \\
\hline
\textbf{R9: Role-based Access Control} & Enforce fine-grained privileges, strict separation of duties, and data-zone isolation to protect sensitive assets. & A9, A18, A19, A21 & Art.~10 (Data Governance - data protection), Art.~15 (Cybersecurity - access controls)\\
\hline
\textbf{R10: Automated Report Generation} & Generate machine- and human-readable technical reports automatically or on demand. & A20, A22, A23 & Art.~11 (Documentation), Art.~12 (Record-keeping - systematic logging) \\
\hline
\textbf{R11: Immutable Audit Trail} & Keep a tamper-evident log of data ingress, code versions, parameters, and results that can be exported to the CA's traceability layer. & A14, A18, A19, A22, A23 & Art.~12 (Record-keeping - comprehensive logs), Art.~15 (Cybersecurity) \\
\hline
\textbf{R12: Scalability \& Resource Efficiency} & Scale elastically for large models or datasets while monitoring compute, storage, and energy. & A18, A27 & Art.~15 (Robustness)\\
\hline
\textbf{R13: Interoperability} & Enable smooth exchange of artefacts and metadata via shared schemas and open formats. & A17, A18, A25–A26, A29 & Art.~11 (Documentation - harmonised formats) \\
\hline
\textbf{R14: Persistent Storage \& Database Management} & Integrate a robust storage and database backend to ensure secure, query-efficient, and long-term management of evaluation data, logs, and artefacts. & A14, A18, A22 & Art.~10 (Data Governance - data management), Art.~12 (Record-keeping - data retention)\\
\hline
\textbf{R15: Cybersecurity and Threat Mitigation} & Apply industry-grade security controls to prevent unauthorised access, detect anomalies, and mitigate risks. & A9, A18, A19, A21 & Art.~10 (Data Governance - data protection), Art.~15 (Cybersecurity - security measures)\\
\hline
\end{longtable}

\subsection{Outlook: Instantiating Customised AITS through a Sandbox Configurator}
\label{sub:instantiation}

Satisfying the requirements of Table~\ref{tab:aits_requirements} one engagement at a time, by hand, is neither scalable for \acp{CA} nor affordable for the SMEs that the AI Act intends to support. What is needed instead is a mechanism that can assemble a compliant testing environment from reusable parts, on the basis of an explicit and reviewable specification of what a given engagement requires. We refer to such a mechanism as a \textit{Sandbox Configurator}.

Rather than a static software component, the Configurator is conceived as a meta-orchestration layer positioned between the regulatory requirements identified above and the concrete tools that implement them. Four design principles follow directly from the requirements. First, sandbox engagements should be described \textit{declaratively}: stakeholders state which evaluation goals, safeguards, and constraints apply, rather than how each is realised, so that the specification remains a readable contract between regulator, provider, and technical expert, and a traceable artefact in its own right (R1, R10, R11). Second, the environment should be assembled from an \textit{external catalogue} of assessment solutions and from a registry of accredited experts able to operate and interpret them, so that methods can be reused and compared across engagements rather than reinvented (R2, R3, R6). Third, extension should occur through \textit{stable plug-in interfaces} and shared schemas, allowing sector-specific tests, visualisers, and data connectors to be added as standards evolve without disturbing the rest of the environment (R7, R13). Fourth, the resulting instance should be \textit{portable}, producing equivalent artefacts and results whether executed on the provider's premises, in a sovereign cloud, or on federated \ac{HPC} nodes such as those offered by the AI Factories, which is what makes evaluations conducted in different jurisdictions mutually intelligible (R8, R12).

An instance assembled in this way carries its infrastructural guarantees with it: role-specific dashboards (R5), low-code pipeline composition \cite{prinz2021low} for users with heterogeneous technical backgrounds (R4), persistent storage (R14), role-based access control (R9), a tamper-evident audit trail (R11), and cybersecurity controls (R15). Because the specification is explicit and versioned, configurations can be revised as a system matures and its risk profile changes, and redeployed without losing traceability or comparability across testing cycles. Reports generated from the sandbox outputs (R10) can be produced in a form compatible with the AIRS exit report, so that technical findings complement the \ac{CA}'s regulatory feedback in a single deliverable.

\begin{figure}[htbp]
\centering
\includegraphics[width=0.8\textwidth]{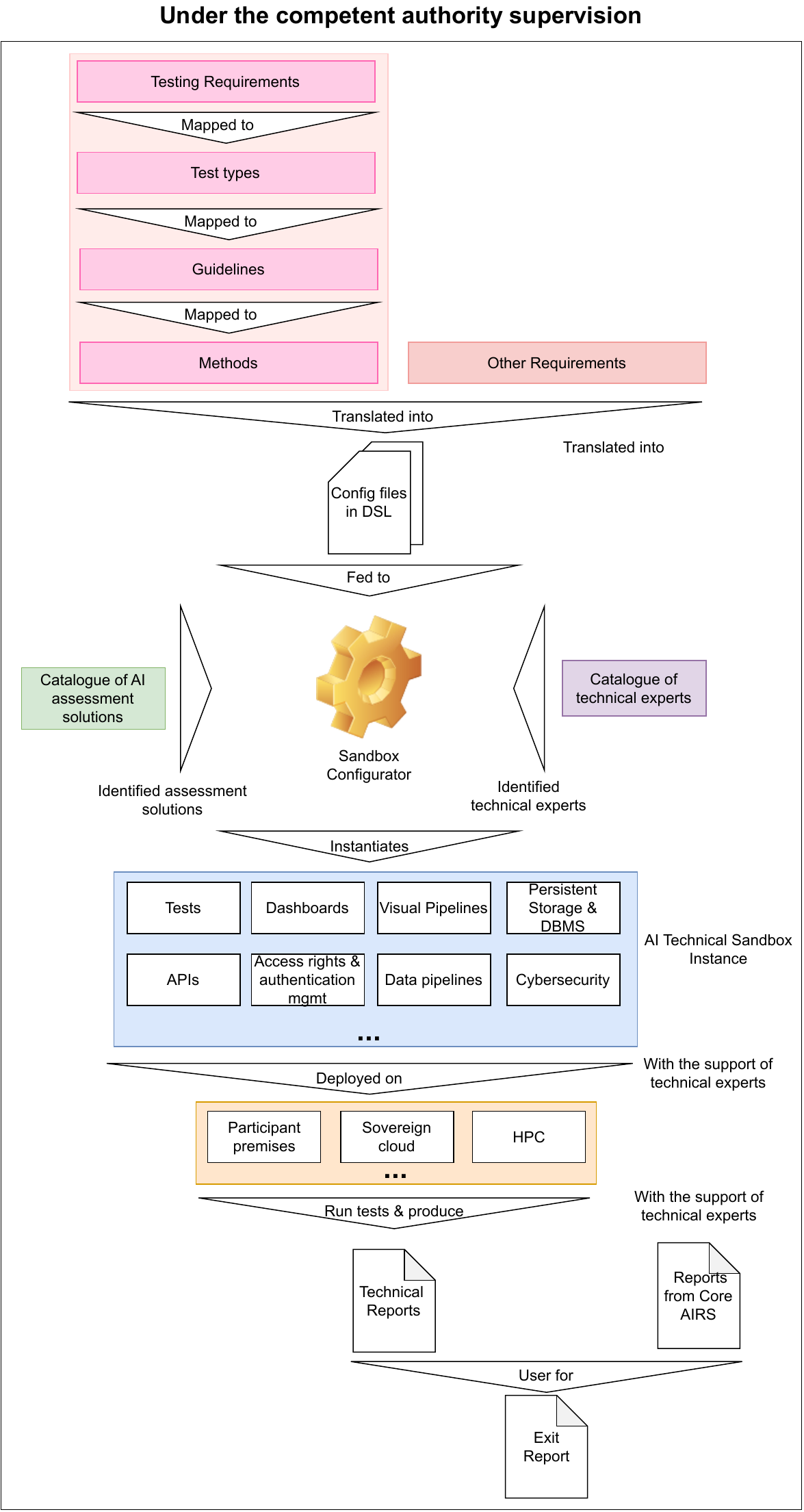}
\caption{High-level overview of the instantiation of an AITS from an explicit sandbox specification.}
\label{fig:example}
\end{figure}

Figure~\ref{fig:example} summarises this logic at a high level. The concrete realisation of the Configurator, including its configuration language, orchestration model, and the tooling on which it is built, is presented in a companion paper \cite{configurator2026}; here we restrict ourselves to the requirements such a framework must satisfy and to the rationale connecting them to the AIRS activities of Section~\ref{sec:activities}.

\section{Examples of AIRS Journey}
\label{sec:examples}

In this section, we present two hypothetical and purely illustrative participant journeys within the AIRS. These scenarios are not based on actual cases but are intended to demonstrate how the activities and requirements identified above play out in practice along both the Core and Extended pathways, and how different EU initiatives may contribute to their realisation. The Core pathway centres on regulatory guidance, whereas the Extended pathway involves more resource-intensive technical testing.

\subsection{Example 1 - Core AIRS}
\label{sub:example1}

ACMEStartup, a young company specialising in affective computing, has developed AffectSense, an AI system for real-time emotion recognition in customer service environments. Designed to operate effectively across multilingual and multicultural contexts, AffectSense is currently being piloted in high-impact applications,such as recruitment support,that fall under the high-risk categories listed in Annex III of the \ac{AI Act}. Given the system's potential to influence individuals' access to essential services, and its reliance on sensitive contextual inference, ACMEStartup learns about the possibility of participating in AIRS to obtain structured regulatory guidance ahead of its product launch.

With support from its local \ac{EDIH}, the company confirms that its use case qualifies as high-risk under the AI Act, due to its potential impact on fundamental rights. As ACMEStartup does not yet have a dedicated Compliance Officer, participation in AIRS represents a valuable opportunity to gain legal clarity, particularly given the cost-free access. Confident in the technical maturity of its system, which has already undergone extensive internal testing, the company opts to participate in Core AIRS, which focuses entirely on regulatory assessment.

With assistance from the CA of its home country, ACMEStartup prepares and submits its application which, after the selection process, is accepted. During the three-month sandbox engagement, the CA responds to the company's regulatory questions and reviews its internal controls.

Minor issues related to documentation and access management are identified and resolved. The CA also provides guidance on the implementation of additional safeguards. Following this phase, ACMEStartup notifies the CA of its intention to proceed with live testing under official supervision. The request is approved, and the company launches a six-month pilot in collaboration with local stakeholders, coordinated through its regional EDIH.

Upon successful completion of the pilot and receipt of final feedback from the CA, ACMEStartup moves forward with the full market deployment of AffectSense.

\subsection{Example 2 – Extended AIRS}
\label{sub:example2}

Safe Corp., a multinational leader in safety and surveillance technologies, has developed an advanced camera system for deployment in public spaces. The system is designed to detect and anticipate hazardous behaviours by pedestrians or vehicles, such as jaywalking or running red lights, at intersections and traffic signals. When a potential risk is identified, the system triggers a real-time alert to help prevent accidents. To ensure compliance with privacy regulations, the system applies anonymisation to general monitoring data, but retains identifiable records strictly when a confirmed violation is detected, so that fines or legal proceedings can be supported in compliance with privacy regulations.

Through its engagement with CitCom.ai,the \ac{TEF} dedicated to smart cities \cite{citcom_ai_2025}, Safe Corp. learns about the Extended \ac{AIRS} and the opportunity to obtain joint regulatory and technical validation. CitCom.ai suggests that the use case qualifies under the high-risk category of the AI Act and offers support in accessing the Extended AIRS stream.

Safe Corp.'s compliance team proceeds with the application to the AIRS, which is accepted by the National CA. The engagement begins with the definition of a technical evaluation strategy, centred on a structured mapping process that translates regulatory expectations into concrete testing requirements. Initial high-level requirements include resilience to input variation, absence of discriminatory behaviour, and clarity of decision-making outputs. These are systematically mapped to appropriate test types, such as robustness evaluation, fairness auditing, and transparency analysis. For each test type, corresponding technical guidelines are selected, drawing from international standards and established best practices. These guidelines are then linked to specific evaluation methods, including benchmark datasets, software tools, metrics, and protocols that can be operationalised within the sandbox environment.

At this stage, the agreed test plan is formalised into an explicit sandbox specification and used to instantiate the technical environment, assembling the required tools, datasets, and benchmarks. The execution and oversight are entrusted to technical experts provided by CitCom.ai, selected from a catalogue of accredited technical experts.

In parallel to the technical configuration, the CA conducts a review of Safe Corp.'s internal controls and governance procedures. This review focuses on how the company manages risk, ensures accountability, and safeguards data protection and transparency in the deployment of its camera system. Minor issues related to internal governance are identified and resolved early in the process, and the CA provides targeted recommendations to strengthen specific compliance measures.

With all inputs consolidated, a portable, policy-compliant sandbox instance is generated. This instance integrates all necessary components: visual pipelines for test composition, real-time dashboards, role-based access controls, persistent storage systems, tamper-evident audit trails, and built-in cybersecurity mechanisms. The sandbox is first deployed within Safe Corp.'s internal infrastructure, where three iterative testing cycles are conducted over six months. Each iteration incorporates technical feedback from the sandbox and regulatory insights from the CA, leading to progressive refinements of both the AI system and the evaluation framework. Because each revision is expressed in the same explicit specification, every reconfiguration maintains full traceability and semantic consistency.

In the eighth month, the final evaluation phase begins. The sandbox instance is deployed on \ac{HPC} infrastructure provided by a national AI Factory, enabling large-scale simulation of operational conditions. Continuity with previous iterations is preserved while the evaluation is scaled up. Test results are visualised through automatically generated dashboards, and a comprehensive technical feedback report is compiled directly from the sandbox outputs. Concurrently, the CA performs a final compliance review, drawing on traceability data, audit logs, and visual artefacts exposed by the sandbox.

The results of both the technical and regulatory evaluations are consolidated into a single Exit Report, which is formally delivered to Safe Corp. This report concludes the sandbox engagement and provides an auditable foundation for any subsequent conformity assessment or market deployment.
In addition, anonymised and non-confidential insights from the experience are shared with the broader European community to foster knowledge transfer.

\section{Discussion}
\label{sec:discussion}

In this section, we reflect on the broader implications of the framework, examining both its scope and limitations as well as its potential to promote harmonisation of sandboxing practices across Europe.

\subsection{Scope and Limits}
\label{sub:scope}

An open-source approach to sandbox tooling raises important considerations regarding its potential use beyond the public regulatory sphere. Although the framework has been primarily conceived to support the implementation of AIRS under the supervision of a \ac{CA}, its modular architecture also lends itself to broader applications. Companies can deploy it internally to conduct self-assessments or embed it within proprietary compliance workflows, thereby enhancing the transparency and consistency of their governance processes. In doing so, they benefit from aligning their internal practices with formats and procedures already recognised by regulators, even if the accountability is not equivalent to supervised sandbox testing. This alignment can facilitate dialogue with authorities and increase trust in company-led risk governance.

Conversely, in the case of Core AIRS, \acp{CA} may rely on results generated outside the sandboxing environment, provided they are grounded in the same methodology. Such results cannot substitute the legitimacy of public AIRS engagements, but they can still provide valuable inputs, reduce duplication of effort, and strengthen transparency between regulated actors and supervisors.

This complementarity highlights a key distinction: while private deployments contribute to internal governance and market assurance, regulatory sandboxes remain irreplaceable for addressing legal grey zones, interpreting obligations in disruptive contexts, and offering systemic oversight that no corporate process can replicate. \acp{CA} alone can translate abstract legal provisions into operational guidance, arbitrate contested issues, and identify emerging systemic risks. These are responsibilities that cannot be delegated to automated pipelines or absorbed into compliance departments.

At the same time, an open-source design enables a shared infrastructure for both regulators and companies across Member States. This fosters methodological exchange, accelerates institutional learning, and gradually harmonises interpretative practices across jurisdictions. Such convergence not only reinforces consistency in supervision, but also ensures that private and public uses of the framework evolve along mutually reinforcing trajectories.

Finally, the effectiveness of any sandbox instance ultimately depends on what is put into it and how the results are interpreted. Tooling can formalise processes, standardise reporting, and provide reusable testing modules, but it cannot determine the right questions to ask or the weight to give to results. Complex AI systems, particularly generative AI, demand contextual expertise to translate technical findings into regulatory or ethical meaning. Standards may offer reference points, but they will not eliminate the need for expert judgement. In this sense, the framework complements rather than replaces the role of regulators and internal compliance: it provides structure and transparency, but the responsibility for framing assessments and interpreting outcomes remains firmly human.

\subsection{Towards Harmonised European Sandboxing Efforts}
\label{sub:harmonisation}

While the primary goal of the proposed framework is to support the structured and scalable implementation of AI Regulatory Sandboxes within individual Member States, its architectural properties also open up compelling opportunities for long-term convergence across Europe.  
A shared technical foundation is envisioned not merely as a deployment framework but as the backbone for a federated ecosystem of interoperable sandboxes. By formalising testing configurations and assembling sandbox instances from modular components, it can promote consistency without imposing uniformity. This allows Member States to tailor sandbox deployments to their unique regulatory, institutional, or sectoral contexts, while still adhering to a shared technical grammar. In turn, practices can be replicated, compared, and iteratively improved across jurisdictions.  

At the time of writing, Member States are exploring different models for deploying AIRS, with some favouring horizontal approaches that span all sectors, and others considering domain-specific sandboxes tailored to particular use cases. The framework is designed to support both paths. Its modular design allows it to be adapted to sectoral constraints while preserving a common structure that facilitates coordination. A harmonised approach across Member States, enabled by shared tools and transparent methodologies, would not only support mutual learning but also ease the consolidation and comparison of findings at the European level.  
This, in turn, enhances the strategic role of the AI Office, which benefits from access to aggregated insights into the evolving AI landscape. By ensuring that sandbox activities remain interoperable and visible, the framework strengthens the Commission's capacity to monitor trends, identify emerging risks, and respond to innovation in a timely and coherent manner.  

Thanks to its modular, open-source, and extensible design, the framework is conceived as a reusable foundation that supports cross-border compatibility without requiring centralisation. National guidelines, sector-specific requirements, and bespoke evaluation tools can be accommodated, while shared test types, templates, and expert modules are reused. This balance between customisation and interoperability is expected to lower the entry barrier for Member States and accelerate the creation of \ac{AIRS}es that are locally adapted yet technically aligned.  
Over time, such alignment could promote greater methodological coherence within the European sandboxing landscape. The adoption of common configuration formats, standardised evaluation workflows, and portable deployment models would reduce fragmentation and enhance comparability and reusability of sandbox outcomes. As stressed in recent work \cite{due2024sandboxing}, knowledge-sharing mechanisms are crucial to avoid duplication, enable regulatory learning, and support harmonisation while still preserving local flexibility. Technically compatible infrastructures thus make coordination both feasible and economically efficient, even before formal harmonisation is mandated at the political level.  

In this context, European initiatives such as \acp{EDIH}, AI Factories, and \acp{TEF} provide fertile ground for scaling this vision. These initiatives already deliver critical capabilities, from early-stage prototyping to high-performance testing, which could be integrated directly into future sandbox deployments. Their distributed presence, sectoral specialisation, and multidisciplinary expertise make them ideal anchors for both national roll-outs and cross-border collaboration. In particular, \acp{EDIH} can provide local support to orient potential participants, helping them to understand the sandboxing process, prepare for applications, and formalise the requirements needed for participation, an element identified in the Section \ref{sub:scope} as fundamental to the success of any sandboxing experience.

TEFs, by contrast, could serve as default providers of technical expertise in their respective domains. Their infrastructure and specialised experience position them to support rigorous evaluations, especially within Extended AIRS. At the European level, CoordinaTEF \cite{ec:tefs} can amplify this role by aligning TEFs, promoting knowledge exchange, and reducing duplication of effort, thereby strengthening their contribution to cross-border sandboxes. Similarly, AI Factories may act as technical anchors in verticals such as healthcare, manufacturing, or cybersecurity. Most importantly, they offer access to AI-dedicated \ac{HPC} infrastructure for secure large-scale technical testing.

By aligning sandbox tooling with these initiatives, Member States could progressively build a federated network of interoperable sandboxes. Such a network would enable the exchange of testing assets, joint oversight of transnational use cases, and the emergence of domain-specific clusters capable of addressing challenges at the European level. Crucially, this vision does not require a centralised authority, but rather a shared commitment to open standards, technical transparency, and cooperative governance.  
While the creation of cross-border sandboxes and a harmonised European sandboxing architecture is not yet an explicit regulatory mandate, the framework is designed to make such outcomes technically attainable. In doing so, it would support compliance with the AI Act while advancing a broader European model of responsible, participatory, and adaptive AI governance, rooted in diversity of implementation and unity of purpose.

\section{Conclusion}
\label{sec:conclusion}

This paper presented a practical and modular framework for implementing \ac{AIRS}es under the EU AI Act, addressing the core challenges of translating legal obligations into testable procedures, supporting iterative improvement, and ensuring interpretability of results. 
Within the \ac{AIRS} defined by the Act, we distinguished between two complementary modes of engagement: a Core AIRS, led by the \ac{CA} to provide regulatory oversight and guidance, and an Extended AIRS, which builds on this foundation by embedding structured technical testing through the AITS. 
This distinction is analytical and participant-oriented, helping to clarify how different needs and levels of readiness can be accommodated without implying multiple sandbox regimes.

To ground this framework, we mapped 29 concrete activities spanning the entire sandbox lifecycle, from onboarding and requirements mapping to testing, feedback, and exit. 
These activities were aligned with 15 infrastructural and governance requirements that we identified as essential for trustworthy AI evaluations, such as transparency, modularity, reproducibility, and auditability. 
For high-risk AI systems, these requirements were explicitly mapped to provider obligations under Articles~9–15 of the AI Act, covering risk management, data governance, documentation, record-keeping, transparency, oversight, robustness, and cybersecurity.
Taken together, the activities and requirements provide a comprehensive reference for both procedural coordination and technical implementation, offering a pathway toward harmonised, interoperable, and innovation-friendly regulatory sandboxes across Europe.

On this basis, we outlined the \textit{Sandbox Configurator}, an open-source framework intended to bridge the gap between high-level regulatory obligations and executable assessments by instantiating tailored sandbox environments from explicit, reviewable specifications. Its design and implementation are reported in a companion paper \cite{configurator2026}.

We highlighted two key dimensions of the framework's role. First, its scope and limits: while it can also be deployed by private actors for internal risk assessments, only public AIRS supervised by \acp{CA} can provide systemic oversight, legal certainty, and legitimacy. In this regard, the framework supports the contextualisation of AI evaluation, which is particularly critical for complex and generative systems where emergent behaviours cannot be anticipated ex ante. Expert judgement across legal, ethical, and technical domains remains indispensable, and the framework's flexibility ensures that testing protocols, metrics, and thresholds can be adapted to specific use cases rather than fixed in absolute terms.
Second, its contribution to European harmonisation: by enabling Member States to instantiate nationally adapted yet technically compatible sandboxes, it can reduce fragmentation and promote methodological convergence. Whether AIRS are adopted in horizontal or sector-specific form, a shared technical foundation supports comparability, peer-to-peer exchange, and timely dissemination of insights to the AI Office. By aligning with EU initiatives such as EDIHs, TEFs, and AI Factories, it also provides a pathway toward a federated, interoperable sandboxing ecosystem rooted in open standards, transparency, and cooperative governance.

As future work, we plan to expand the mapping between testing requirements and evaluation methods, and to validate the activities and requirements presented here against real sandbox engagements. Within the \ac{L-AIF} \cite{luxai_factory}, LIST and the \ac{SnT} (University of Luxembourg) have already initiated this effort. In parallel, through the CitCom.ai TEF, we have established an "AI Assessment Club", in which we are working to identify legal, ethical, and technical requirements, together with standards and methodologies relevant for AI adoption in Smart Cities and Communities. In addition, through the \ac{RE.M.I.} initiative \cite{remi}, a collaboration between the \ac{L-AIF} and the \ac{CNPD}, Luxembourg's market surveillance authority under the AI Act, we are promoting a community of practice that defines requirements and establishes feedback loops with the market. This ensures continuous alignment between regulatory priorities and industrial practices, while also developing a national catalogue of experts and supporting Luxembourg's national strategy on AI \cite{LuxembourgAI2025}. 

To progress from national pilots toward a shared European framework, we invite Member States, regulatory authorities, research institutions, and industry stakeholders to contribute to the development and implementation of a shared sandboxing infrastructure. As a first step, interested parties are encouraged to reach out to the authors of this white paper to initiate dialogue and explore concrete opportunities for collaboration. Through pooled resources, aligned practices, and open dialogue, Europe can establish a robust, interoperable, and future-ready foundation for trustworthy AI experimentation and oversight.

\section*{Acknowledgements}
This project was partially supported by CitCom.ai, co-funded by EU/Digital Europe and, in Luxembourg, by the Feder; by Luxembourg AI Factory, co-funded by EU/Digital Europe and, in Luxembourg, by the Government.

\bibliographystyle{IEEEtran}
\bibliography{references.bib}

\end{document}